\documentclass{emulateapj}

\lefthead{van der Wel et al.}
\righthead{Size and Number Density Evolution of Early-Type Galaxies}
\slugcomment{Accepted April 8, 2009}

\newcommand{\kms}{\>{\rm km}\,{\rm s}^{-1}}
\newcommand{\reff}{R_{\rm{eff}}}
\newcommand{\msol}{M_{\odot}}

\begin{document}

\title{On the Size and Comoving Mass Density Evolution of Early-Type
  Galaxies}

\author{Arjen van der Wel, Eric F. Bell, Frank C. van den Bosch, Anna
  Gallazzi \& Hans-Walter Rix} \affil{Max-Planck Institute for
  Astronomy, K\"onigstuhl 17, D-69117, Heidelberg, Germany;
  vdwel@mpia.de}

\begin{abstract}
  We present a simple, empirically motivated model that simultaneously
  predicts the evolution of the mean size and the comoving mass
  density of massive ($>10^{11}~\msol$) early-type galaxies from $z=2$
  to the present.  First we demonstrate that some size evolution of
  the population can be expected simply due to the continuous
  emergence of early-type galaxies.  The Sloan Digital Sky Survey
  (SDSS) data reveal that in the present-day universe more compact
  early-type galaxies with a given dynamical mass have older stellar
  populations. This implies that with increasing look-back time, the
  more extended galaxies will be more and more absent from the
  population.  In contrast, at a given stellar velocity dispersion,
  SDSS data show that there is no relation between size and age, which
  implies that the velocity dispersion can be used to estimate the
  epoch at which galaxies stopped forming stars, turning into
  early-type galaxies.  Based on this, we define an empirically
  motivated, redshift-dependent velocity dispersion threshold above
  which galaxies do not form stars at a significant rate, which we
  associate with the transformation into early-type galaxies.
  Applying this 'formation' criterion to a large sample of nearby
  early-type galaxies, we predict the redshift evolution in the size
  distribution and the comoving mass density.  The resulting evolution
  in the mean size is roughly half of the observed evolution.  Then we
  include a prescription for the merger histories of galaxies between
  the 'formation' redshift and the present, based on cosmological
  simulations of the assembly of dark matter halos.  Such mergers
  after the transformation into an early-type galaxy are presumably
  dissipationless ('dry'), where the increase in size is expected to
  be approximately proportional to the increase in mass.  This model
  successfully reproduces the observed evolution since $z\sim 2$ in
  the mean size and in the comoving mass density of early-type
  galaxies with mass $M>10^{11}~\msol$.  We conclude that the recently
  measured, substantial size evolution of early-type galaxies can be
  explained by the combined effect of the continuous emergence of
  galaxies as early types and their subsequent growth through dry
  merging.
\end{abstract}


\section{INTRODUCTION}
Over the past five years it has become clear that early-type galaxies
do not constitute a passively evolving population. Their predominantly
old stellar populations notwithstanding, substantial evolution occurs
up until the present day. The main evidence for such evolution is the
increase in comoving mass density by a factor of about three from
$z\sim 1$ to the present
\citep[e.g.,][]{bell04b,brown07,faber07}. Recently, additional
evidence for continuing structural evolution of the early-type galaxy
population was provided by the observation that distant ($z>1.5$),
quiescent galaxies have much smaller sizes than early-type galaxies of
the same stellar mass in the present-day universe
\citep{daddi05,trujillo06b,zirm07,toft07,vandokkum08b,cimatti08,buitrago08}.
This surprising result, raising questions about the evolutionary
connection of these compact galaxies with present-day descendants, was
recently put on firmer footing by \citet{vanderwel08c}: using
early-type galaxies with accurate dynamical masses, thus removing the
most important systematic problem that may hamper the $z\sim 2$
results (the absolute uncertainty in the mass estimates), they
inferred substantive size evolution, by a factor of $\sim 2$, at a
given mass, between $z=1$ and present.  Recently, also
\citet{bernardi09} showed that size evolution continues up until the
present day, and must be a smooth function of redshift.  Although the
remaining uncertainties are not negligible \citep[see,
e.g.,][]{hopkins09b}, it has become clear that the observed size
evolution cannot be explained by systematic uncertainties.

Because halos that collapsed at earlier epochs were denser, and
gas-rich, dissipative processes were more important for galaxy
formation, it has been argued that galaxies that formed early are
smaller \citep[e.g.,][]{robertson06,khochfar06b}.  In this scenario,
because the number of early-type galaxies increases with cosmic time,
and the more recently formed galaxies are larger, there is a direct
connection between the size and comoving mass density evolution in the
early-type galaxy population. This is true even without individual
galaxies changing in size over time.

This paper explores the extent to which the growth of the population
can account for the observed size evolution, and/or whether additional
mechanisms to increase the size evolution of individual galaxies are
required.  As a proof of concept, we first investigate the present-day
early-type galaxy population for clues that indicate that evolution in
the mass-size relation may be expected purely due to evolution of the
population, not size growth of individual galaxies.  Are small/compact
early-type galaxies older in terms of their stellar populations than
large/extended early-type galaxies with the same mass?  In Section
\ref{sec:model1} we show that this is indeed the case.  Is there a
parameter for which age and size are independent?  Continuing along
the track of previous work
\citep[e.g.,][]{bernardi05a,chang06,graves08}, which demonstrated that
stellar velocity dispersion ($\sigma_*$) is the fundamental parameter
behind well-known correlations such as the color-magnitude and
mass-metallicity relations, we show that $\sigma_*$ also correlates
with, and perhaps drives, the age of the stellar population.  This
suggests that formation epoch and $\sigma_*$ are related.

We cast this idea as an empirical relationship between the velocity
dispersion of an early-type galaxy and the redshift $z_{\rm{ET}}$ at
which it emerged as an early-type galaxy.  This model to describe the
population evolution of early-type galaxies allows us to predict which
portion of the present-day early-type galaxy population already
existed at a given redshift $z$.  In turn, such a
$\sigma_*-z_{\rm{ET}}$ relationship in conjunction with the
present-day $\sigma_*$ distribution thus describes the evolution with
redshift of the properties of the population as a whole, in particular
its size distribution.  This is akin to the concept of 'progenitor
bias' \citep{vandokkumfranx01}.  As it turns out, a significant size
evolution is expected even in the absence of size evolution of
individual galaxies.

Then we continue and include a prescription for subsequent merging
based on the cosmological simulations of the dark matter halo assembly
by \citet{li07}.  The merger rates that we derive are in agreement
with those found empirically \citep[e.g.,][]{bell06,lotz08,lin08}.
Because we examine the merger history after the last, major episode of
star-formation activity and the transformation into an early-type
galaxy, by definition, these mergers are dissipationless or 'dry'
\citep{vandokkum05,bell06,naab06}.  Under reasonable assumptions for
the relationship between the properties of progenitors and descendants
based on numerical simulations of merging, gas-poor progenitors
\citep[e.g.,][]{ciotti01,gonzalez03,nipoti03,boylan05,robertson06}, we
can predict how individual galaxies grow. This evolution of individual
galaxies is then superimposed on the evolution of the population
(Section \ref{sec:model2}).  A comparison with the observations allows
us to investigate whether the continuous growth of the early-type
galaxy population, with galaxies experiencing 'puffing' through dry
mergers, is able to explain the observed size evolution with redshift,
or whether other, perhaps unknown, physical mechanisms must play a
significant role.  Potential caveats, known problems, and testable
predictions of our model are discussed in Sections
\ref{sec:discussion} and \ref{sec:discussion:4}.  We adopt as
cosmological parameters
$(\Omega_{\rm{M}},~\Omega_{\Lambda},~h,\sigma_8) =
(0.3,~0.7,~0.7,~0.9)$.

\section{Summary of High-Redshift Observations}\label{sec:data}
Before describing and testing our model, we summarize the relevant
high-redshift measurements that we seek to explain. These are the
evolutions of the characteristic size and the comoving mass density of
early-type galaxies from $z=1$ and $z=2$ to the present.

\subsection{The Early-Type Galaxy Population at $z\sim1$}\label{sec:data:1}
Size evolution between $z\sim 1$ and the present has recently been
measured by \citet{vanderwel08c}, who found, at a given dynamical
mass, $\reff(z)\propto(1+z)^{-0.98\pm0.11}$ for galaxies with masses
$M>10^{11}~\msol$.  This corresponds to
$\delta\reff(1)\equiv\reff(1)/\reff(0.06)=0.54\pm0.04$ ($z=0.06$ is
the center of the redshift bin in which we select our sample of nearby
early-type galaxies described in Section \ref{sec:model1:sample} and
will be used throughout as a benchmark).

The second observational quantity we compare our model with is the
evolution of the comoving mass density $\rho$. We derive the change in
the comoving mass density of early-type galaxies with masses larger
than $10^{11}~\msol$ from the evolution of the normalization
($\phi^*$) of the luminosity function for red galaxies as measured by
\citet{faber07}. Even though not all massive, red galaxies have
early-type morphologies, the fraction of early-type galaxies among
these changes little between $z=1$ and the present
\citep[e.g.,][]{bell04a,vanderwel07b}.  We assume that the
characteristic mass at the 'knee' of the mass function $M^*=7.1\times
10^{10}~\msol$ does not change, implying that the evolution in
$\phi^*$ is identical to that in $\rho$.  This assumption should hold
because the evolution of the $M/L$, as derived from evolution of the
fundamental plane zero point and the observed evolution of $L^*$
cancel.  However, it cannot be ruled out at this stage that $M^*$
changes by $\sim0.1$ dex.  From the results presented by
\citet{faber07}, we derive that
$\delta\rho(1)\equiv\rho(1)/\rho(0.06)=0.35\pm0.13$ for galaxies with
mass $M>10^{11}~\msol$, where the error includes an uncertainty of 0.1
dex in the evolution of $M^*$.

\begin{figure*}[t]
  \epsscale{.35}
  \plotone{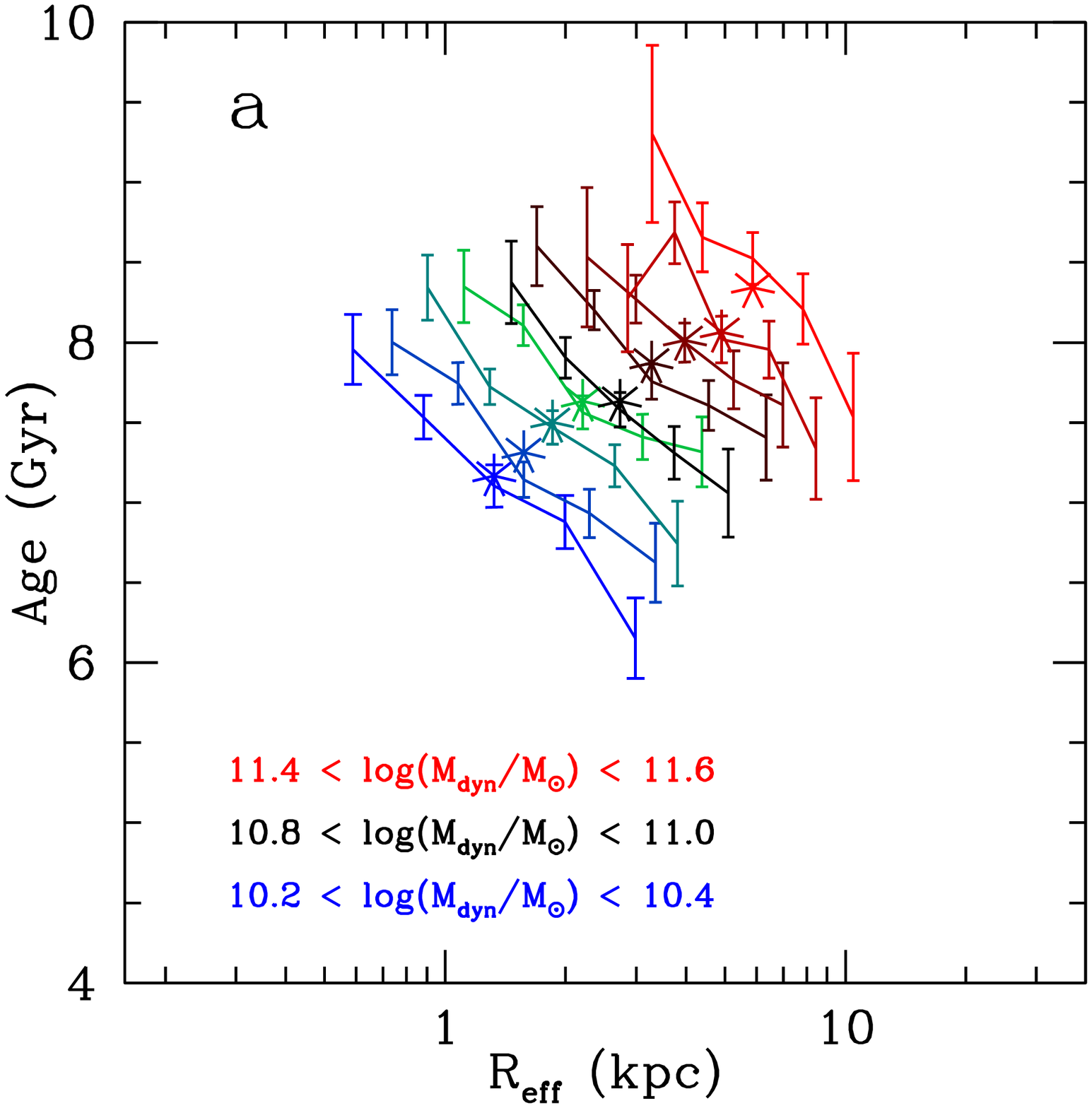}\hspace{0.4cm}\plotone{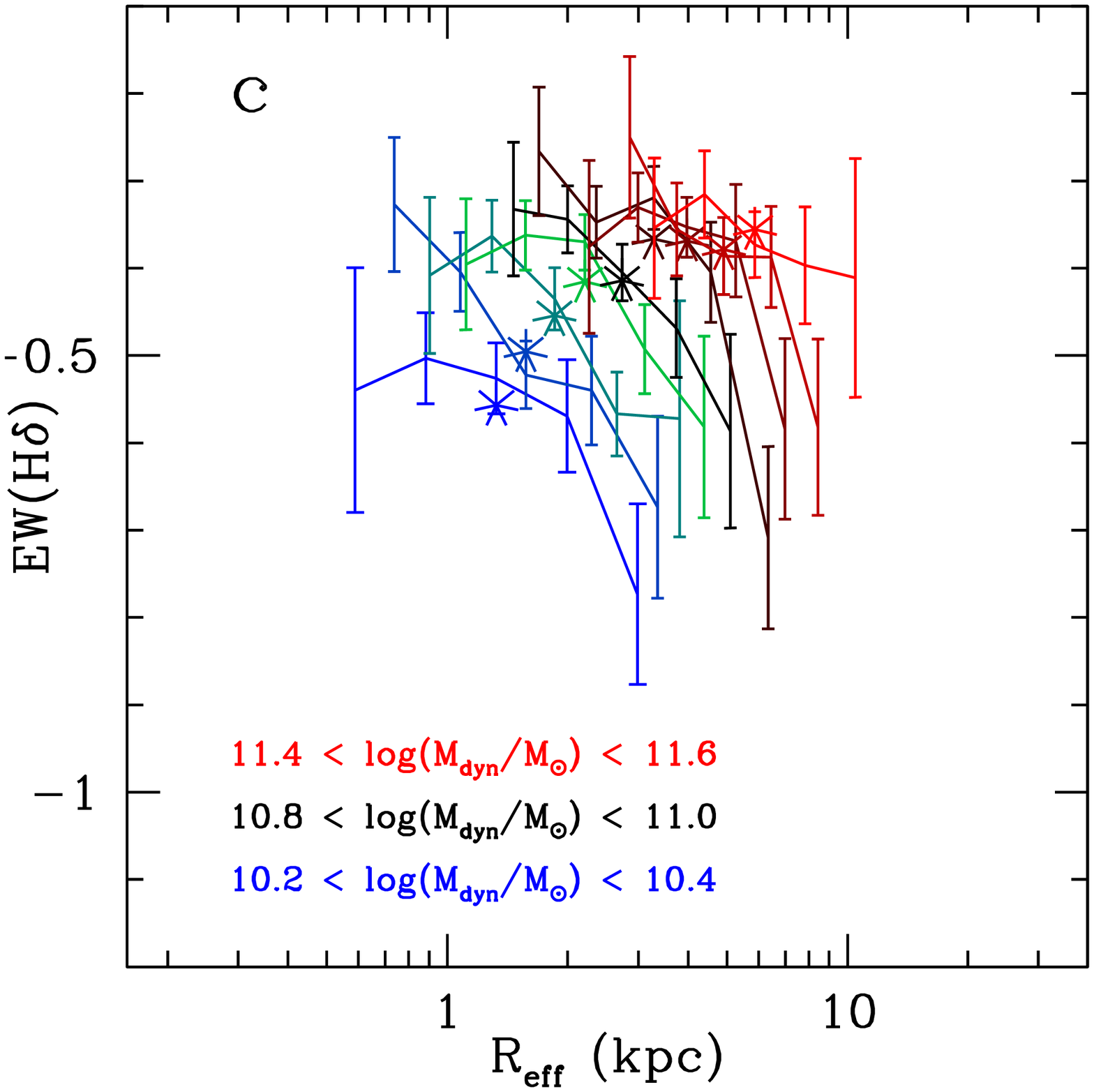}\hspace{0.4cm}\plotone{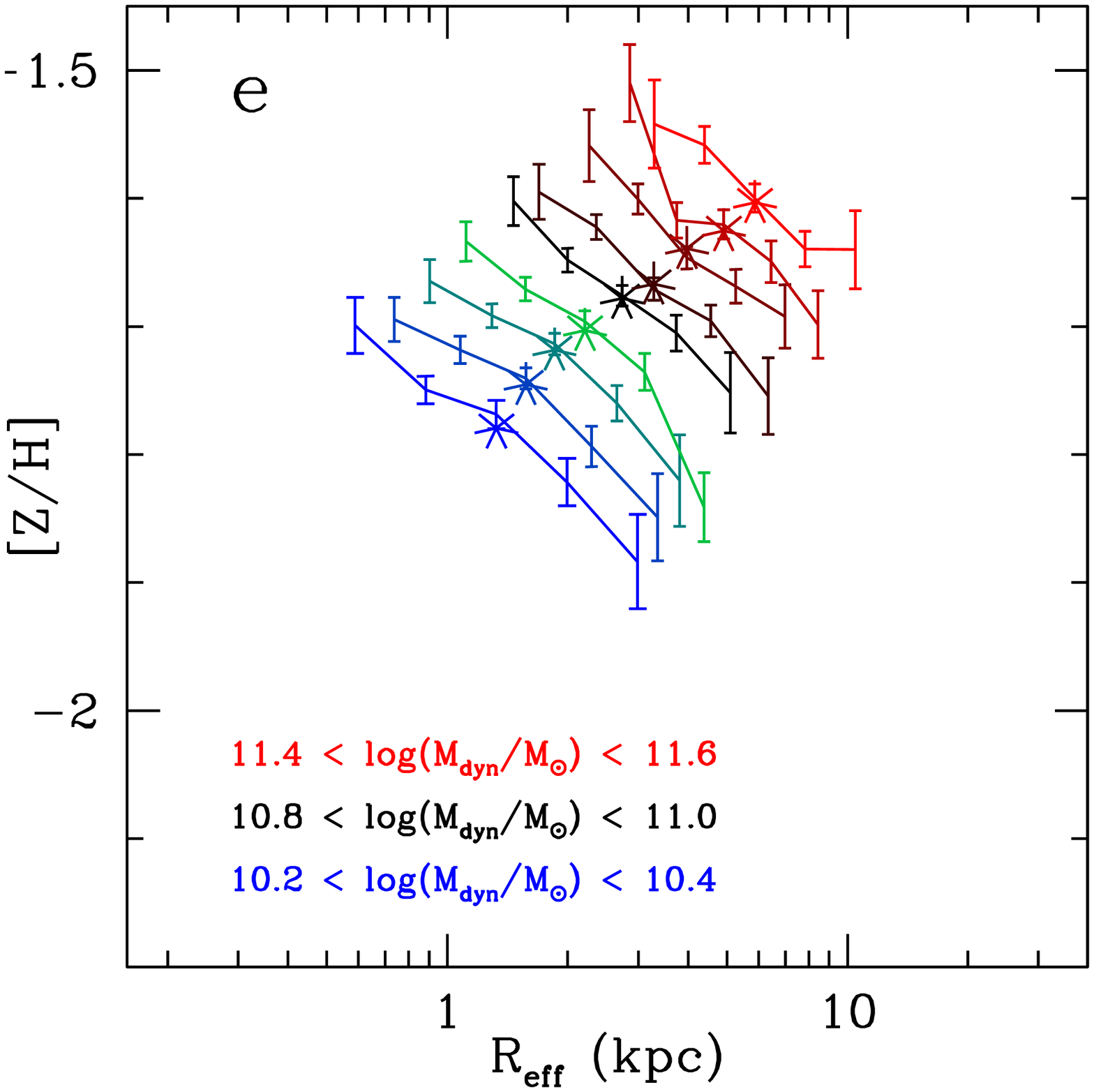}
  \plotone{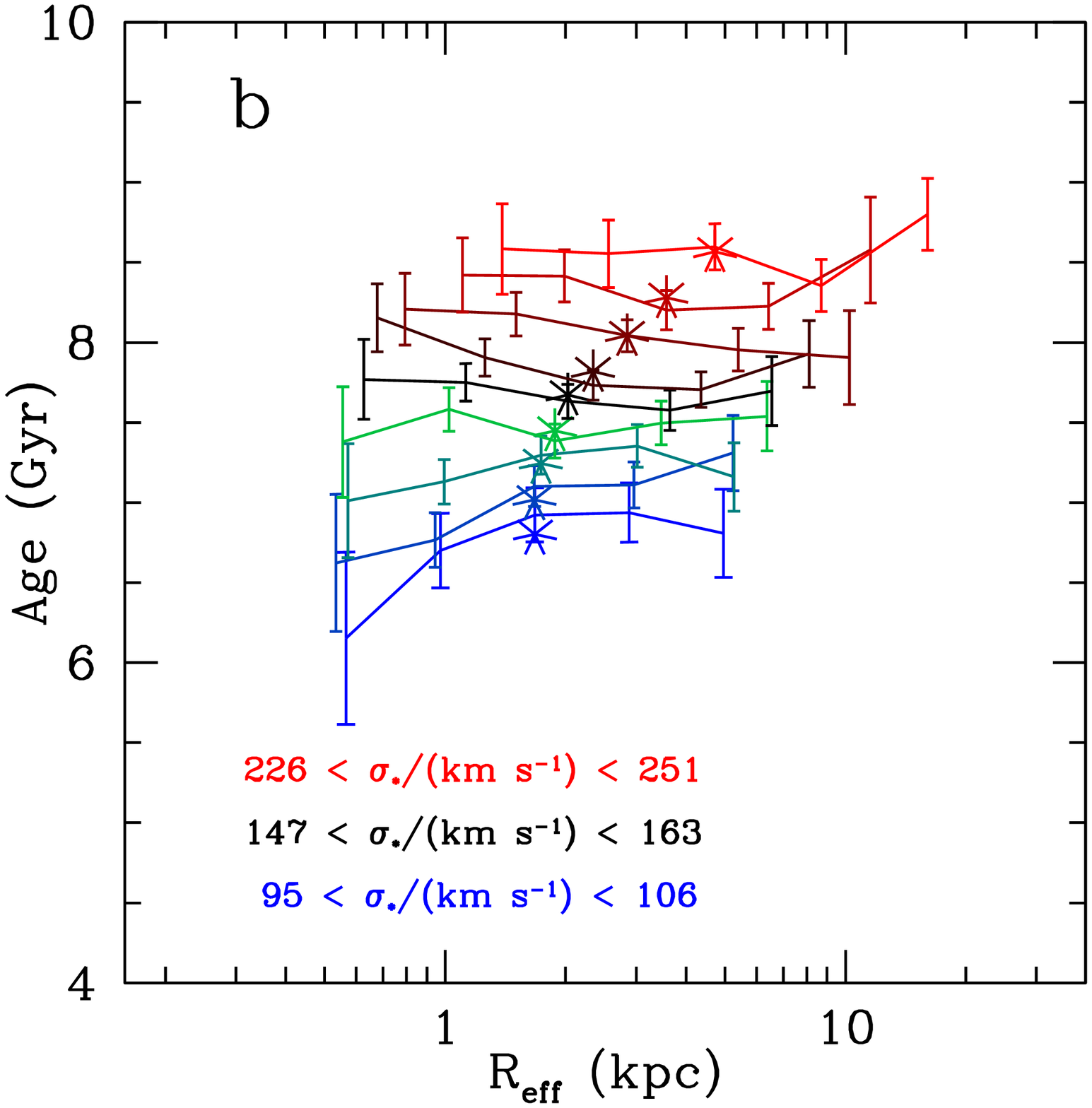}\hspace{0.4cm}\plotone{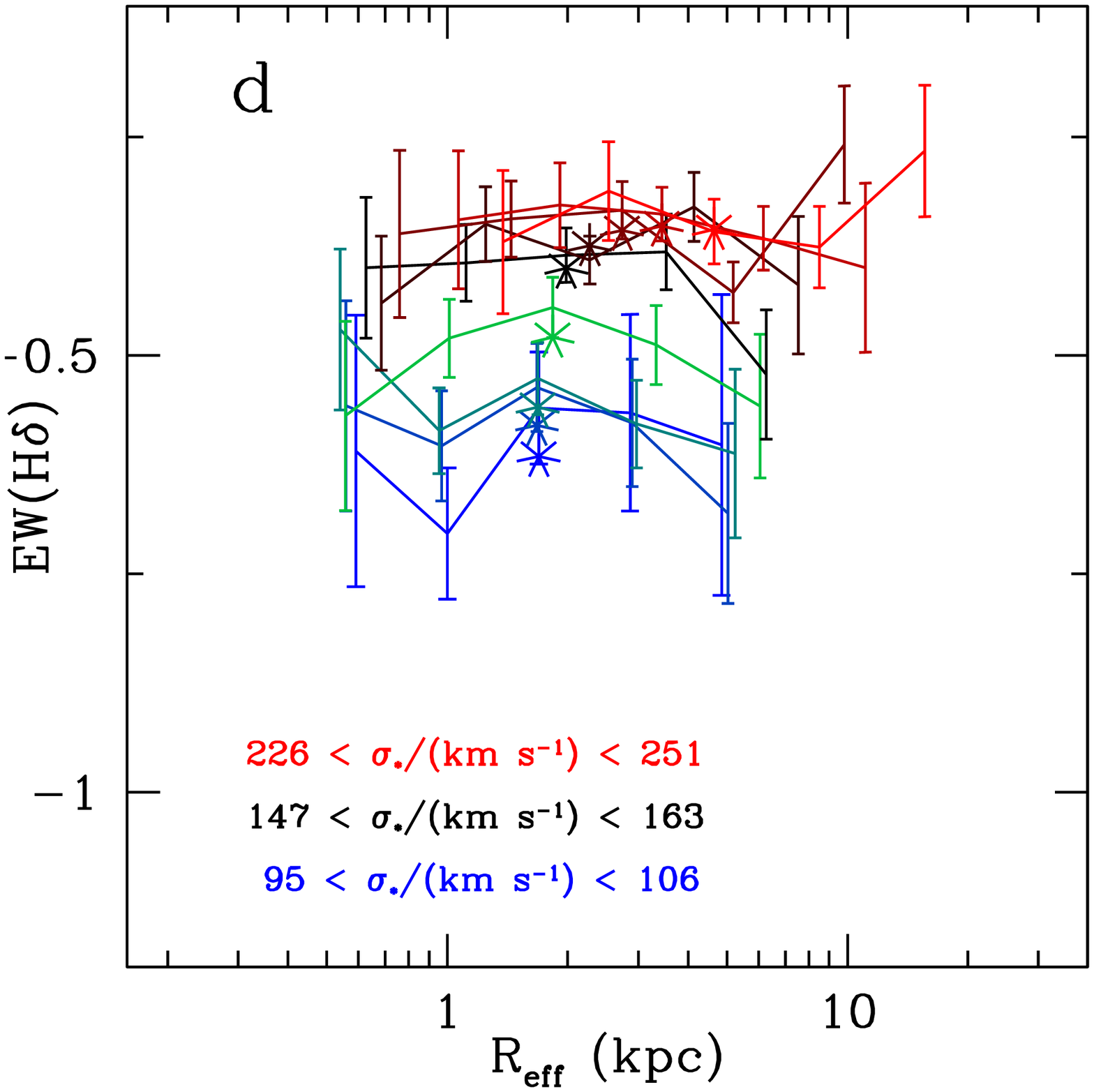}\hspace{0.4cm}\plotone{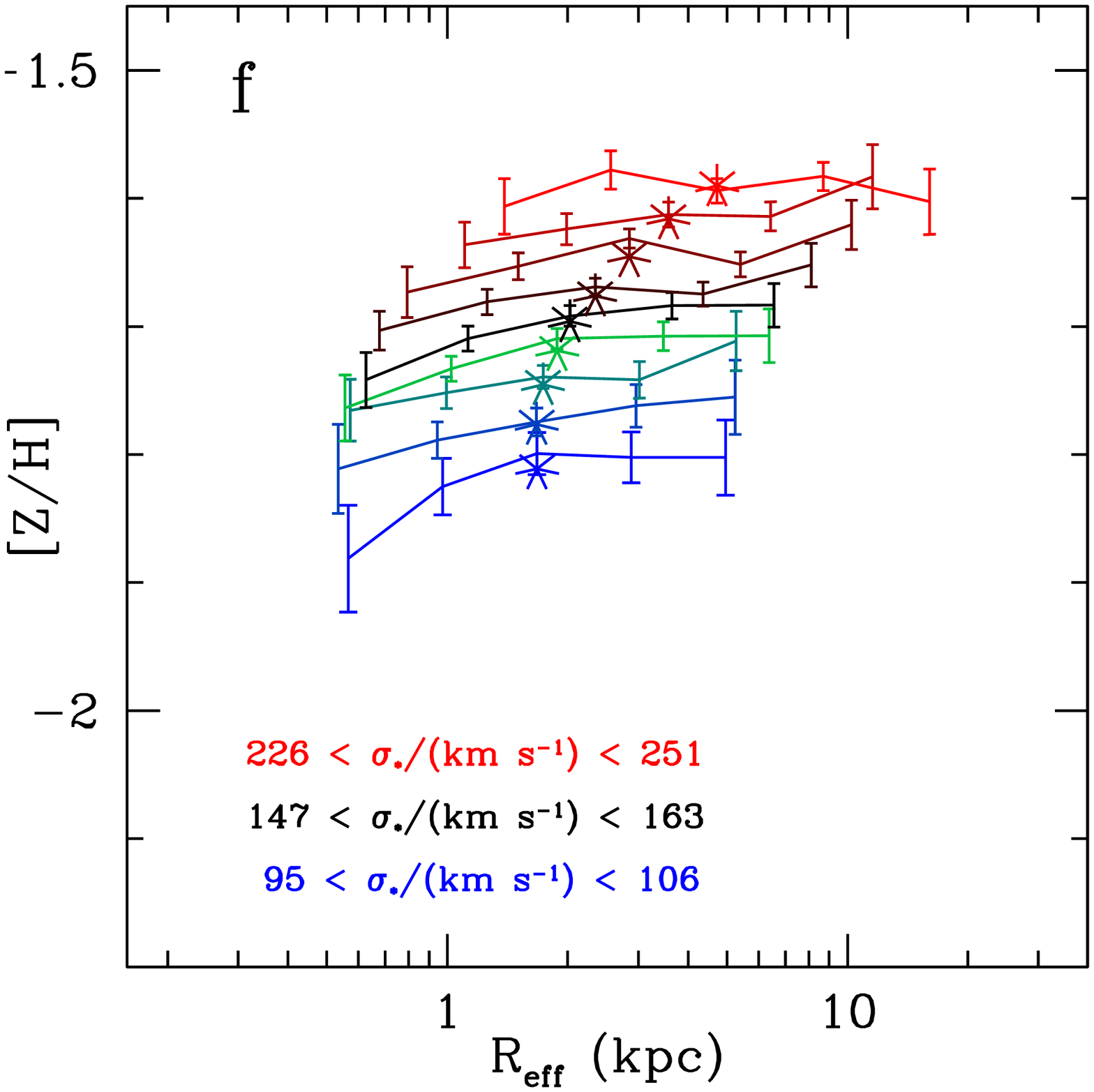}
  \caption{\textit{(a)}: The age-size relation for present-day
    early-type galaxies in nine different dynamical mass bins. The age
    is the light-weighted stellar population age. The distance between
    the same-colored error bars is equal to the scatter (1$\sigma$) in
    size.  The median size and age in each bin, indicated by the
    stars, both increase with mass (from blue to red), reflecting the
    well-known mass-age and mass-size relations.  However, at fixed
    mass small galaxies are older than large galaxies. \textit{(b)}:
    The age-size relation as in panel \textit{(a)}, but here in nine
    bins of stellar velocity dispersion instead of $M_{\rm{dyn}}$. The
    same overall trend is seen in the sense that galaxies with high
    $\sigma_*$ are, on average, larger and older than galaxies with
    low $\sigma_*$. However, at \textit{fixed} $\sigma_*$ the age-size
    trend differs importantly from that seen at fixed $M_{\rm{dyn}}$:
    age does not depend on size (or size does not depend on age) at
    fixed $\sigma_*$. \textit{(c)} and \textit{(d)}: The same as
    panels \textit{(a)} and \textit{(b)}, respectively, but with
    H$\delta$ line strength instead of age along the y-axis.  The
    trends are similar as those seen with age, albeit with larger
    uncertainties. Since H$\delta$ is a robust age indicator, the
    similar behavior in this figure of H$\delta$ and age demonstrates
    that the trends seen with age are not caused by model
    uncertainties in the age estimates, and cannot be
    artificial. \textit{(e)} and \textit{(f)}: The same as panels
    \textit{(a)} and \textit{(b)}, respectively, but with metallicity
    instead of age along the y-axis.  As with age, there is no
    correlation between size and metallicity at fixed $\sigma_*$.  We
    note that the size of the SDSS spectroscopic aperture cannot
    explain the observed trends (see Section \ref{sec:model1:prop} for
    details).}
\label{R_AGE}
\end{figure*}

\subsection{The Early-Type Galaxy Population at $z\sim2$}
\label{sec:data:2}
The observational constraints at higher redshifts are far less secure
than at $z\lesssim 1$ due to limited sample sizes and systematic
uncertainties.  Simply extrapolating the $z=1$ size evolution
measurement mentioned above to $z=2$, without regard to other
measurements, gives $\delta\reff(2)=0.36\pm0.04$.  Compiling all
available data sets at higher redshifts (up to $z=2.5$),
\citet{vanderwel08c} found marginally faster evolution of
$\reff(z)\propto(1+z)^{-1.20\pm0.12}$, i.e.,
$\delta\reff(2)=0.29\pm0.04$. Assuming that the difference between the
robust, but generously extrapolated, $z\sim 1$ results, and the
direct, but systematically uncertain, $z\sim 2$ results is indicative
of the true, systematic error, we conservatively adopt
$\delta\reff(2)=0.3\pm0.1$ as the most realistic estimate of the size
evolution between $z=2$ and the present.  We note that all
measurements have been obtained for galaxies with masses
$M>10^{11}~\msol$. At lower masses samples are non-existent or
severely biased.

The observational constraints on $\phi^*$ and hence $\rho$ are less
secure at $z\sim 2$ than at $z\sim 1$, again because of systematic
uncertainties. The latest estimate by \citet{marchesini08} yields
$\delta\rho(2.4)=0.14^{+0.05}_{-0.08}$ for all galaxies with stellar
masses larger than $10^{11}~\msol$\footnote{The mass estimates in that
  study assume the \citet{salpeter55} initial mass function for which,
  given the other uncertainties, the difference between stellar mass
  and dynamical mass is negligible for the mass range of interest here
  \citep{vanderwel06b,borch06,cappellari06}.}.  Other recent
determinations \citep{elsner08,perez08} are consistent with this
estimate. \citet{kriek08b} have shown that already at $z\sim 2.4$ the
majority ($56\%^{+8\%}_{-12\%}$) of such massive galaxies have red
colors, and that those show no or little evidence for star-formation
activity. This number is somewhat lower than for the nearby massive
galaxy population which has a red sequence fraction of 77\%.  In
addition, despite the quiescent nature of the red $z\sim 2$ galaxies,
it is as yet unclear what their morphologies are.  They are barely
spatially resolved, such that it is impossible to tell whether they
have smooth surface brightness profiles, and constraints on the shape
of their surface brightness profiles are rather poor, even though
\citet{toft07} showed that exponential profiles are perhaps a better
representation than De Vaucouleurs profiles.  Given these constraints
on the evolution of the red/early-type fraction among the massive
galaxies at $z\sim 2$, we arrive at
$\delta\rho(2.4)=0.10^{+0.04}_{-0.06}$ as our best estimate of the
comoving mass density of early-type galaxies at $z=2.4$ with mass
$M>10^{11}~\msol$.

\section{A Simple Empirical Model I: \\ Density Reflects Formation
  Epoch}
\label{sec:model1}
\label{sec:model1:sample}
We develop an empirical estimate of the formation epoch of early-type
galaxies based on the correlations between the global properties of
the present-day population of early-type galaxies.  In this approach,
individual early-type galaxies are assumed not to evolve after their
formation in either mass or size.  A sample of 17,483 nearby
early-type galaxies at redshifts $0.04<z<0.08$ has been constructed by
\citet{graves08} from the Sloan Digital Sky Survey (SDSS) database
\citep[DR6,][]{adelman08}.  Only galaxies with measured velocity
dispersions, on the red sequence, either without emission lines or
with high $[\rm{OII}]$-to-$\rm{H}\alpha$ ratios and with concentration
parameters $C>2.5$ are included in the sample.  These criteria
effectively exclude star-forming galaxies, but include genuine
early-type galaxies with nuclear activity \citep[see][]{yan06}. For
the description of the velocity dispersions and the determination of
effective radii, as well as our dynamical mass estimator
($M_{\rm{dyn}}\propto R\sigma_*^2$), we refer to \citet{vanderwel08c}.
In addition, we require that luminosity-weighted stellar ages are
available from \citet{gallazzi05}; 16,279 out of 17,483 galaxies have
such an age estimate.

\begin{figure}
\epsscale{1}
\plotone{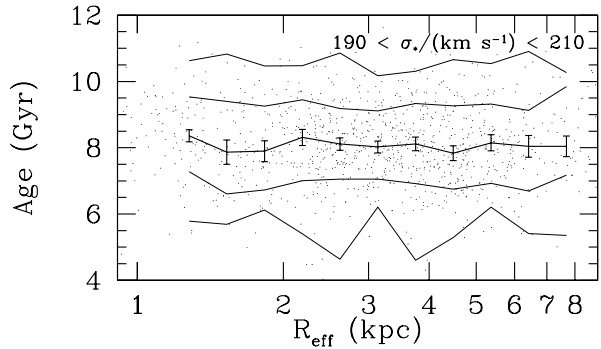}
\caption{ Size vs. stellar population age for present-day early-type
  galaxies in a narrow range of velocity dispersion. The line with the
  errorbars is a running median, while the other lines indicate the
  scatter ($1\sigma$ and $2\sigma$). The scatter is of order 10\% and
  the residual from the running median does not correlate with any
  other parameter.}
\label{R_AGE1}
\end{figure}

\subsection{Correlations between Global Properties of Early-Type
  Galaxies}
\label{sec:model1:prop}
It is well known that early-type galaxies with large masses and
velocity dispersions have older stellar populations than those with
low masses and velocity dispersions
\citep{trager00b,thomas05,gallazzi06,graves07}.  In Figure
\ref{R_AGE}a we also show that in our SDSS sample massive galaxies
have older stellar populations.  In the same figure, it can also be
seen that, naturally, size increases with mass.  As a corollary, we
have that, loosely speaking, galaxies with large sizes have older
stellar populations than galaxies with small sizes.

For these broad, global trends it does not matter what tracer of mass
is adopted. Dynamical mass and velocity dispersion both show similar
correlations with size and age (compare Figures \ref{R_AGE}a and
\ref{R_AGE}b).  However, the difference between the dynamical mass and
the velocity dispersion comes to light when we dissect the sample, and
look at the relation between age and size at fixed dynamical mass or
at fixed velocity dispersion.  At a fixed dynamical mass, there is an
anti-correlation between size and age: large galaxies have younger
stellar populations than small galaxies (Figure \ref{R_AGE}a).  This
trend persists over a mass range of almost two orders of magnitude,
from $\sim 10^{10}~\msol$ to $\sim 10^{12}~\msol$, and there is no
clear indication that the slope of the age-size relation depends on
mass.

The correlation between age and size at fixed mass implies that the
zero point and scatter of the mass-size relation will evolve with
redshift, even if we assume that individual galaxies do not change in
size.  This is simply due to the fact that larger, younger galaxies
did not exist yet at earlier epochs.  This is, in essence, the model
prediction made by \citet{khochfar06b}.  The result shown in Figure
\ref{R_AGE}a provide the first empirical evidence for this process.

It is perhaps surprising that the picture changes quite dramatically
if we look for an age-size relation for galaxies with a given velocity
dispersion.  In Figure \ref{R_AGE}b, it can be seen that at a given
velocity dispersion there is no evidence that small galaxies are older
than large galaxies. On the other hand, if one realizes that
$M_{\rm{dyn}}\propto \reff \sigma_*^2$, the size-age relation has to
be flatter at fixed $\sigma_*$ than at fixed $M_{\rm{dyn}}$ (or may
even be reversed).  The fact that the relation should disappear
altogether at fixed $\sigma_*$ is, however, not obvious.  In Figure
\ref{R_AGE1}, we show for one particular narrow bin around
$\sigma_*=200~\kms$ the size-age distribution. Again, there is no sign
that age depends on size (or vice versa). Moreover, the scatter is
remarkably uniform and rather small ($\sim10\%$).  We could not
identify a parameter that correlates with the residual in the size-age
relation at fixed $\sigma_*$. This includes environment: using the
\citet{yang07} SDSS group catalog, we find that the scatter in age
does not correlate with group membership, the mass of the group, the
distance to the group center, or whether galaxies are centrals or
satellites \citep[see also,][]{vandenbosch08b}.

The age estimates are unavoidably model dependent, and may therefore
be hampered by systematic uncertainties.  In Figure \ref{R_AGE}c and
\ref{R_AGE}d we replace stellar population age by H$\delta$ absorption
line strength.  The strength of this Balmer line is relatively
insensitive to metallicity and is therefore a good tracer of age.
Although the random uncertainties are substantially larger, the
picture remains the same: at fixed dynamical mass the H$\delta$ line
is stronger for larger galaxies, implying a younger stellar
population, whereas at fixed velocity dispersion there is no
correlation between size and line strength.  We conclude that the
trends with stellar population age in Figure \ref{R_AGE}a and
\ref{R_AGE}b are robust and not artificial.

In Figures \ref{R_AGE}e and \ref{R_AGE}f, we show the correlation
between size and metallicity.  As with age, for the population as a
whole there is a positive correlation between size and metallicity, as
expected from the combination of the mass-metallicity relation and
mass-size relations.  At fixed velocity dispersion (Figure
\ref{R_AGE}f), however, there is no correlation between size and
metallicity, which is consistent with the idea that the velocity
dispersion traces the potential well depth. By necessity, we also find
that at fixed dynamical mass, the metallicity is smaller for large
galaxies (see Figure \ref{R_AGE}e).

The SDSS spectroscopic aperture does not sample the entire galaxy;
therefore, some aperture effects may be expected: age and metallicity
gradients could introduce artificial trends in Figure \ref{R_AGE}.
However, aperture effects would cause a positive correlation between
age/metallicity and size, not the observed anticorrelation.
Therefore, we conclude that aperture effects do not strongly affect
our analysis.

The emerging picture is consistent with previous work.  It has been
shown that at fixed $\sigma_*$ there is no color-magnitude relation
\citep{bernardi05a,graves08}, which indicates that velocity
dispersion, more so than luminosity, determines the properties of the
stellar population, unless some sort of conspiracy renders trends
invisible.  However, this possibility is excluded by the lack of a
trend in Figures \ref{R_AGE}b, \ref{R_AGE}d, and \ref{R_AGE}f.  In
addition, \citet{chang06} find that, at fixed velocity dispersion,
there is almost no correlation between stellar mass and absorption
line indices.

Very recently, several authors have noted similar trends (or lack
thereof) as in our Figure 1. \citet{graves09} show that, at fixed
velocity dispersion, age and metallicity do not correlate with galaxy
size (see our Figures \ref{R_AGE}b and \ref{R_AGE}f).  Moreover,
\citet{shankar09} derive the same trend that we show in Figure 1a: at
fixed mass, old galaxies are smaller than young galaxies.

Furthermore, \citet{bell00} and \citet{kauffmann04} showed that for
any type of galaxy, stellar mass surface density is a better predictor
than stellar mass of the star-formation history and current
star-formation activity, and surface density is closely related to
velocity dispersion.  This was made explicit by \citet{kauffmann06},
who showed that there is a sharp transition in the fraction of
quiescent versus star-forming galaxies at particular values of both
surface density and velocity dispersion.

These previous results and our result that age and size at a given
$\sigma_*$ are uncorrelated (Figure \ref{R_AGE}b) tell us something
about the significance of $\sigma_*$ as a (and perhaps \textit{the})
fundamental characteristic of a galaxy.  We now pursue the idea that
$\sigma_*$ can be used as a predictor of the epoch when star formation
was truncated and an early-type galaxy emerged.

\subsection{The Formation Epoch of Early-Type Galaxies}
\label{sec:model1:form}
As a starting point, we take the observation by \citet{kauffmann06}
that at the present epoch the transition between star-forming and
quiescent galaxies takes place at $\sigma_*\sim125\kms$.  Truncation
of star formation may be associated with the morphological
transformation to an early-type galaxy, and so we may suppose that
newly formed early-type galaxies in the present-day universe have a
velocity dispersion of $\sigma_{\rm{ET}}=125\kms$.  This transition is
not infinitely sharp; in reality there are late-type galaxies with
higher $\sigma_*$ and early-type galaxies with lower $\sigma_*$. We
ignore this scatter in our model, but in Section
\ref{sec:discussion:1} we will briefly comment on the consequences for
our model predictions.

Given such a velocity dispersion threshold for star-formation activity
in the present-day universe, we may expect that such a threshold
existed at earlier epochs as well, but not necessarily with the same
value. The relation between age and velocity dispersion (Figure
\ref{R_AGE}b) implies that $\sigma_{\rm{ET}}$ would have had to be
higher in the past.  Observational evidence for the existence of
$\sigma_{\rm{ET}}$ at high redshifts was recently found by
\citet{franx08}, who used data extending to $z\sim 3$ to show that
there exists a surface mass density threshold, $\Sigma_{\rm{ET}}$,
above which star-formation activity drops suddenly and significantly.
Interestingly, this threshold was shown to be larger at higher
redshifts.  We parametrize the velocity dispersion threshold for the
truncation of star formation and the transformation into an early-type
galaxy as follows:

\begin{equation}
\label{sigz}
  \sigma_{\rm{ET}}(z) = 125\kms~(1+z)^{\alpha}.
\end{equation}

Consequently, given the velocity dispersion of a galaxy we can
estimate the redshift at which its star formation was truncated and it
attained its early-type morphology:

\begin{equation}
\label{zsig}
  z_{\rm{ET}}(\sigma_*) = \left(\frac{\sigma_*}{125\kms}\right)^{1/\alpha}-1.
\end{equation}

Although precise constraints on the amount of evolution are still
lacking due to systematic errors and small sample sizes, the results
from \citet{franx08} imply that $\alpha$ must lie in the range
$1/4\lesssim\alpha\lesssim3/4$.  In the following, we will adopt
$\alpha=3/4$ as the standard value, which turns out to provide the
best results. We will discuss the dependence of our results on the
chosen value and the range allowed by the data in Section
\ref{sec:discussion}.  We note that the redshift dependence of
$\sigma_{\rm{ET}}$ is consistent with the long-standing result from
fundamental plane studies at high redshift that high-mass galaxies
have stellar populations with high formation redshifts
\citep[$z\gtrsim 2$, see][and references therein]{vandokkum07}, and
also with the observation that this formation redshift depends on
mass, low-mass galaxies having younger stellar populations epochs
\citep{vanderwel05, treu05b}. In addition, this is in agreement with
archaeological studies of nearby early-type galaxies
\citep[e.g.,][]{thomas05,gallazzi08}.

\begin{figure}
\epsscale{1}
\plotone{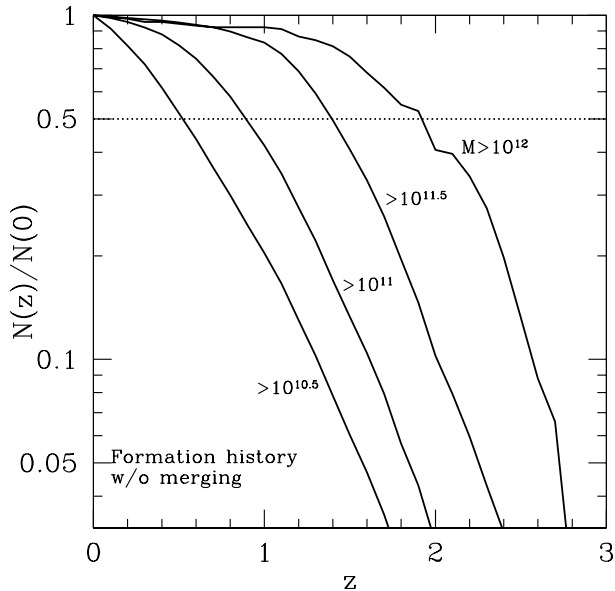}
\caption{Predicted evolution with redshift of the comoving number
  density of early-type galaxies with different masses based on the
  velocity dispersion threshold specified in Equation \ref{zsig}. The
  horizontal dotted line indicates a fraction of 50\%. High-mass
  galaxies emerged as early-type galaxies earlier than low-mass
  galaxies. Note that this evolutionary picture does not include
  merging, as described in Section \ref{sec:model2}.}
\label{evolplot2}
\end{figure}

To quantify the model predictions for earlier epochs, we apply
Equation \ref{sigz} to our nearby galaxy sample, only retaining those
with $\sigma_*>\sigma_{\rm{ET}}(z)$ as galaxies that were early types
already at redshift $z$.  In Figure \ref{evolplot2} we show how the
early-type galaxy population would build up at different present-day
masses for $\alpha=3/4$.  Half of the galaxies with present-day masses
exceeding $10^{12}~\msol$ had become early types by $z\sim 2$, while
half of those with masses exceeding $10^{11}~\msol$ had become early
types by $z\sim 1$.  This behavior is reminiscent of the predictions
by, e.g., \citet{delucia06} and \citet{neistein06}, who also predict
that massive galaxies formed their stars earlier than less massive
galaxies.

\begin{figure*}
\epsscale{.95}
\plotone{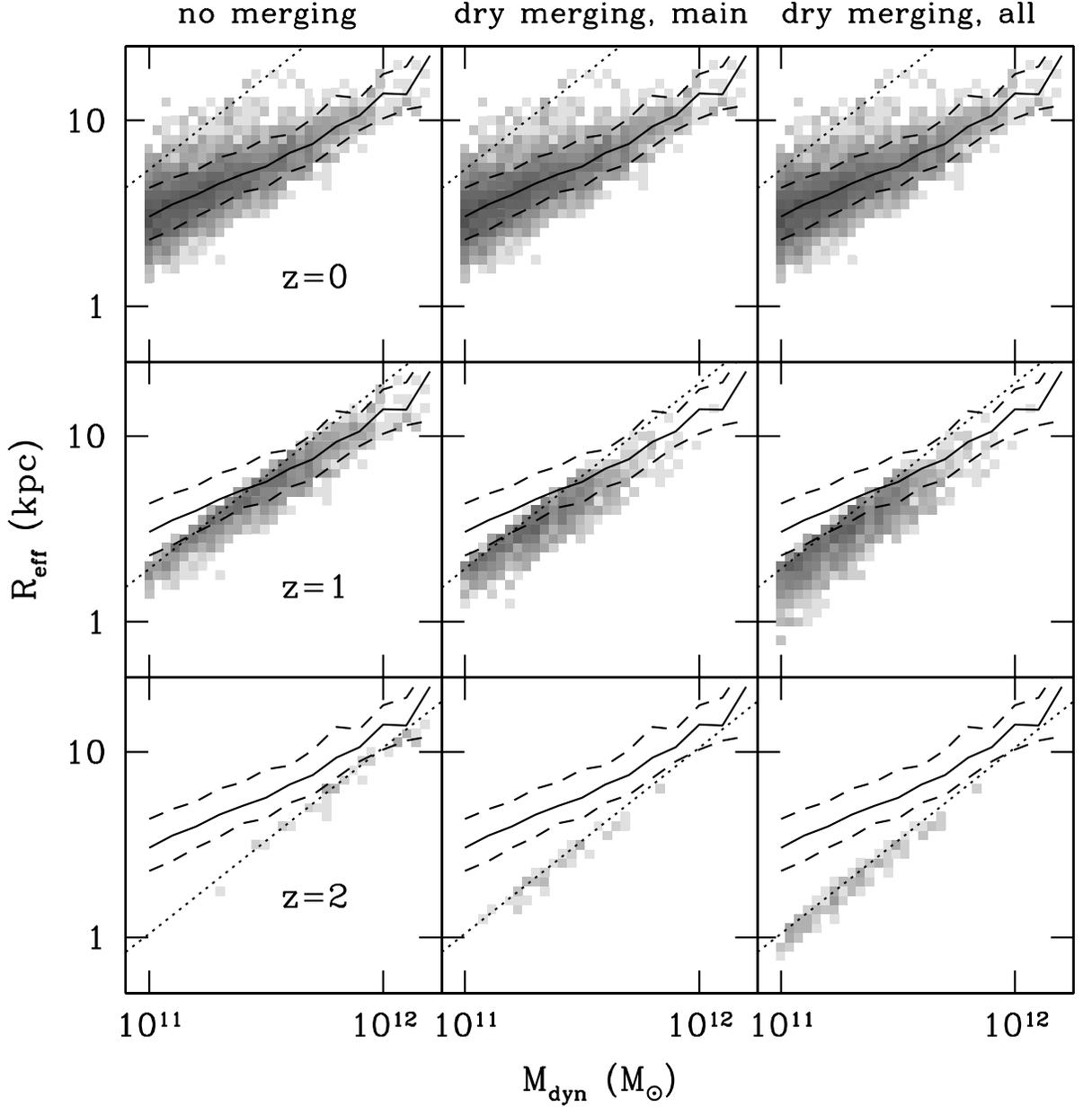}
\caption{Predicted redshift evolution of the mass-size relation. The
  left-hand column of panels shows the predicted evolution for the
  model described in Section \ref{sec:model1:form}, i.e., without
  merging.  The model presented in the middle column includes merging
  as described in Section \ref{sec:model2:main}, only considering the
  main progenitors of present-day early-type galaxies. The model
  presented in the right-hand column includes, in addition, all
  progenitors, as described in Section \ref{sec:model2:accr}.  The top
  row (in which all panels are identical) shows the observed, local
  mass-size relation, which is the starting point of our model, the
  gray-scale corresponding to the number of galaxies.  The solid line
  is a running median, and the dashed lines indicate the scatter
  (1$\sigma$). These lines are repeated in every panel to guide the
  eye.  The diagonal dotted line corresponds to $\sigma_*=125~\kms$,
  which is the present-day velocity dispersion threshold,
  $\sigma_{\rm{ET}}$, for the formation of early-type galaxies
  according to Equation \ref{sigz}.  The middle row shows the
  predicted mass-size relation for $z=1$.  The dotted line now
  corresponds to $\sigma_*=210~\kms$, which is $\sigma_{\rm{ET}}$ at
  $z=1$.  The lighter gray scale reflects the evolution in the number
  of galaxies.  In addition, their size distribution is different.
  The bottom row shows the predicted mass-size evolution for $z=2$,
  where $\sigma_{\rm{ET}}=285~\kms$. In our model, the effect of dry
  mergers is that galaxies move parallel to the dotted lines, i.e.,
  $\sigma$ remains constant (see Section \ref{sec:model2}). }
\label{MR}
\end{figure*}

In addition to comoving mass density evolution, Equation \ref{zsig}
also implies size evolution at fixed dynamical mass: at a given mass,
large galaxies have smaller velocity dispersions and therefore became
early-type galaxies more recently.  Figure \ref{MR} shows the
predicted evolution in the size distribution from the present to $z=1$
and $z=2$.  The behavior of the model is straightforward: at
progressively higher redshifts the minimum velocity dispersion of a
galaxy (the dotted lines in Figure \ref{MR}) is higher, as defined by
Equation \ref{sigz}.  The resulting evolution in both the number of
galaxies and their size distribution can readily be seen.  How does
this model compare with the observations described in Section
\ref{sec:data}? For $z=1$, we find $\delta\rho(1)=0.61$ and
$\delta\reff(1)=0.81$, both for galaxies with mass $M>10^{11}~\msol$,
and where the latter is the average offset from the local mass-size
relation, computed in log-space.  The actually observed evolution is
faster (see Section \ref{sec:data:1} and Figure \ref{evolplot}),
significantly for the evolution in average size and marginally for the
mass density evolution.  For $z=2$ the model predicts
$\delta\rho(2)=0.09$ and $\delta\reff(2)=0.71$. The comoving mass
density in this case agrees well with the observed value. The observed
size evolution, however, is much stronger than this model predicts
(see Section \ref{sec:data:2} and Figure \ref{evolplot}).

Despite the quantitative disagreements, it is encouraging that this
model, which only prescribes a redshift-dependent $\sigma$-threshold
when galaxies stop forming stars and become early types, implies
trends in the right direction.  These results suggest that the
continuous 'top-down' emergence of new early-type galaxies can explain
roughly half of the observed size evolution, with individual galaxies
not changing after their initial transformation.  However, an
indication that individual galaxies must change as well was recently
found by \citet{trujillo09}: the local abundance of galaxies with the
same properties as the compact, high-redshift objects is very low,
implying that such galaxies did not survive in their original form.
It is clear that size evolution of individual galaxies has to occur as
well.

\section{A Simple Empirical Model II: \\ Growth through Mergers}
\label{sec:model2}
We now augment the simplistic model from Section \ref{sec:model1} by a
prescription to include subsequent merger activity of galaxies after
the epoch of transformation into early types.  After constructing
merger histories, we describe how mergers affect, in our model, the
masses, sizes, and velocity dispersions of early-type galaxies.

\subsection{Merger Histories}\label{sec:model2:history}

After they become early types, galaxies still undergo subsequent
evolution in terms of merging and/or minor episodes of star-formation
activity, without changing their overall morphological properties,
apart from, perhaps, brief periods of time after accretion
events. That this can lead to substantial size evolution has been
shown by \citet{naab07}.  To include such evolution schematically into
our model we implement merger histories based on simulations of dark
matter halo assembly from \citet{li07}.

Following \citet{li07}, we define $n$ as the inverse of the fractional
mass increase in a merger. In other words, if the mass of the main
progenitor is $M_i$, then the mass of the merger remnant is
$M_f=(1+1/n)M_i$.  The \citet{li07} simulations demonstrate that the
number of mergers does not depend on $n$, i.e., the probability
distribution of $n$ is flat.  To realize merger histories we can
therefore assume a series of accretion events with randomly chosen
$n$.  We verify this simplified approach by attempting to reproduce
the merger histories that \citet{li07} inferred, starting with seed
halos that have 1\% of their present-day mass.  When we allow mergers
in the range $1<n<100$, we construct an ensemble of merger histories
with the same number of mergers with $n<3$, $n<4$, and $n<6$ as found
by \citet{li07}; see their Figures 11 and 12.  In addition, our
scatter in the number of such mergers is also close to that found by
\citet{li07}. We conclude that it is sensible to approximate the
merger histories of dark matter halos by assuming a series of merger
events with randomly chosen $n$ in the range $1<n<100$.

Since we are not interested in the merger history of a halo growing
from 1\% of its present-day mass, but rather in the merger history of
a halo between $z=1$ or $z=2$ and the present, we need to know by how
much halos grow over that period of time. Contrary to the distribution
of $n$, the growth with time depends on halo mass
\citep{vandenbosch02,wechsler02}.  We focus on halo masses in the mass
range $M_{\rm{halo}}=10^{12}-10^{13}\msol$, which corresponds to the
halo masses of galaxies with masses in the range $M\sim
10^{11}-10^{12}\msol$.  In the \citet{li07} simulations, halos in this
mass range had assembled $\sim50\%$ of their present-day mass by $z=1$
($M_{z=1}=M_{z=0}^{-0.30\pm 0.09}$, where 0.09 is the scatter), and
$\sim30\%$ by $z=2$ ($M_{z=2}=M_{z=0}^{-0.55\pm 0.28}$).

Combining the elements discussed so far, we have the following
practical method to construct the possible merger history between
$z=1$ and the present for a halo with present-day mass $M$.  We assume
a series of mergers with mass ratios $1/n_i$, with $1<n_i<100$ a
sequence of randomly chosen real numbers. For example, the most recent
merger involved a main progenitor with mass $n_1M/(n_1+1)$ and a minor
(or, accreted) progenitor with mass $M/(n_1+1)$.  The main progenitor
is, in turn, assumed to be the product of a merger with ratio
$1/n_2$. The main progenitor is followed in this manner, assuming a
sequence of mergers with ratio $1/n_i$, up until the point that the
expected mass for the main progenitor at $z=1$ is reached.  This
expected mass is $M_{z=1}=M_{z=0}^x$, where $x$ is drawn from a
Gaussian distribution with a mean $-0.30$ and a standard deviation
$0.09$. To construct the merger history between $z=2$ and the present,
the only difference is that we choose $x$ from a Gaussian distribution
with mean $-0.55$ and a standard deviation $0.28$.  In practice, in
order to conserve mass, the mass ratio of the final event is rounded
such that the $M_{z=1}$ is precisely $M_{z=0}^x$.  In case $x>0$,
which we formally allow, we assume that no mergers have occurred.

This process describes the merger history of a halo.
\citet{vandenbosch08a} found that the vast majority of the galaxies in
the mass range of interest here should be central galaxies.  We,
therefore, assume that the merger history of the halo directly
corresponds to the merger history of the galaxy.  We incorporate one
exception to this rule into the model: mergers involving halos with
mass $M_{\rm{halo}}<10^{11}~\msol$ do not increase the mass of the
galaxy, but only the mass of the halo. This takes into account that
such small halos contain a much smaller fraction of stars
\citep[e.g.,][]{cooray02,vandenbosch07}.  The growth in galaxy mass is
therefore somewhat slower than the growth in halo mass. As it turns
out, our model predictions are not strongly affected by this. In
Section \ref{sec:discussion:2}, we discuss the effect of possible
delays between mergers between halos and mergers between their
occupying galaxies.

For each galaxy in the nearby sample, we construct a Monte Carlo
realization of its merger history as described above, which predicts
by how much the mass of the main galaxy grows between $z=1$ ($z=2$)
and the present, and how the accreted mass is distributed over a
number of additional progenitors at $z=1$ ($z=2$). Hence, we have two
sets of merger trees: one that describes the $z=1$ progenitors and
another that describes the $z=2$ progenitors.  The average number of
mergers per galaxy with mass ratios exceeding 1:2 is $\sim0.25$
between $z=1$ and the present ($\sim0.55$ between $z=2$ and the
present); the average number of mergers with mass ratios exceeding 1:4
is $\sim0.40$ ($\sim0.90$).  Note that whereas the dry merger rate per
unit time \textit{per early-type galaxy} was higher in the past, the
dry merger rate per unit time and \textit{per unit comoving volume}
was smaller due to the strong evolution in the comoving density of
early-type galaxies.  The merger rates in our model are consistent
with or lower than the available observational constraints on the dry
merger rate between $z\sim 1$ and the present
\citep{bell06,lotz08,lin08}.

\subsection{The Effect of Dry Mergers on Galaxy Sizes}
\label{sec:model2:mergesize}
A crucial part of the model is the effect of merging on the sizes of
galaxies.  Because we are interested in the merger history of galaxies
that are already early types, we focus on dissipationless (dry)
merging.  According to most numerical simulations the remnants of
gas-poor mergers of equal-mass progenitors situated in dark halos have
roughly twice the size of those progenitors
\citep[e.g.,][]{ciotti01,gonzalez03,nipoti03,robertson06}; yet
\citet{boylan05} find that the size of the remnant is typically only
1.5 times that of the progenitors.  This difference may, in some
cases, be explained by the assumed orbits: the bound orbit used by
Boylan-Kolchin et al. may be responsible for remnants that are smaller
than found by, e.g., Ciotti et al., who use a parabolic orbit.
However, not all simulations with bound orbits lead to small remnants
\citep[e.g.,][]{robertson06}, such that the reason for the discrepancy
is not completely clear.  In agreement with the majority of the
predictions, we assume that equal-mass, dry merger products have
double the size of the progenitors (see Section \ref{sec:discussion}
for further discussion).  The virial theorem then implies that the
velocity dispersion remains constant, supported by the numerical
simulations.  Moreover, the results of \citet{gonzalez03} and
\citet{nipoti03} suggest that for a dry merger with any mass ratio the
size of the main progenitor increases linearly with mass, keeping its
velocity dispersion constant.  The implied decrease in surface
brightness is such that mergers of this kind move galaxies roughly,
but not precisely, along the fundamental plane \citep[see,
e.g.][]{robertson06}.  This is reassuring because scenarios in which
this is not the case are difficult to reconcile with the observed
tightness of this scaling relation \citep[e.g.,][]{rix94}.  These
considerations allow us to directly and linearly link the growth in
mass and size in our merger scenario.

\subsection{The Evolution of the Main Progenitors}
\label{sec:model2:main}
First, we investigate the evolution of the main progenitors in the
merger scenario described above, ignoring the smaller progenitors.
Because we assume that mergers do not alter the velocity dispersion of
the main progenitor, it is straightforward to combine the predicted
merger history from Section \ref{sec:model2:history} with the
formation redshift criterion described in Section
\ref{sec:model1:form}.  Figure \ref{MR} shows how the evolution in the
number of early-type galaxies and their size distribution is altered
when we augment the $\sigma_{\rm{ET}}$ model from Section
\ref{sec:model1} by dry merging, only considering the main
progenitors.  Relative to the scenario without merging, the number of
high-mass galaxies decreases more rapidly with redshift, while the
number of lower-mass galaxies decreases less rapidly. The average
difference in size is predicted to be somewhat larger than in the
absence of merging.  The agreement with the observations improves
considerably (see Figure \ref{evolplot}): The merger model now
predicts a redshift evolution of $\delta\reff(1)=0.68$ and
$\delta\reff(2)=0.45$ for the average size at a given mass, and
$\delta\rho(1)=0.38$ and $\delta\rho(2)=0.03$ for the comoving mass
density.  The agreement with the observations is not perfect though;
the observed evolution of the mean size is still $\sim30\%$ faster
than predicted.

\subsection{The Properties of Accreted Progenitors}
\label{sec:model2:accr}
At redshifts $z\lesssim 1$, when few major mergers occur and the
growth of halos is relatively slow, it is probably sufficient to
consider only the main progenitors, as we did in the previous
section. However, at higher redshifts, when major mergers are more
frequent and halos grow fast, this simplification may become
inappropriate.  It is, therefore, also necessary to include the
smaller progenitors in the analysis and assess the question how their
masses and sizes affect the evolution of the population averages. In
the following we do not consider merger activity of the smaller
progenitors before their accretion onto a larger object; thus, some of
the uncertainty remains.

We make two assumptions regarding the properties of the minor
progenitors. First, because we are mainly interested in dry merging,
we assume that all progenitors are early-type galaxies. Second, we
assume that the velocity dispersions of all progenitors are equal to
the dispersion of the main progenitor, and therefore also equal to
that of the final descendant (see Section \ref{sec:model2:mergesize}).
The second assumption is seemingly quite extreme and is obviously not
valid for minor mergers. However, since the goal is to predict only
the abundance and sizes of progenitors with mass $M>10^{11}~\msol$,
this is not necessarily a problem. Moreover, Equation \ref{sigz}
implies that, in our model, all early-type galaxies at any epoch have
a minimum velocity dispersion, $\sigma_{\rm{ET}}$. At redshifts $z\sim
2$, where smaller progenitors matter, the number of early-type
galaxies is a rapidly declining function of velocity dispersion such
that most galaxies available for dry merging will have comparable
velocity dispersions.  Thus, given the characteristics of our model so
far, it is defensible that galaxies that partake in mergers at $z\sim
2$ with mass ratios 1:3 or 1:4 have similar velocity dispersions.

\begin{figure}
\epsscale{1}
\plotone{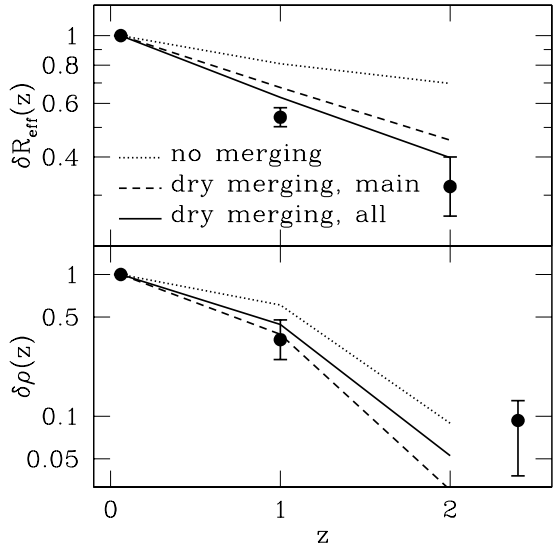}
\caption{ Comparison of the observations described in Section
  \ref{sec:data} (points with error bars) with the model predictions
  for redshift evolution of the average size and fixed mass (the top
  panel) and the total, comoving mass density (the bottom panel) for
  all early-type galaxies with mass $M(z)>10^{11}~\msol$. The dotted,
  dashed, and solid lines refer to different versions of our model, as
  in Figure \ref{MR}.  The dotted line represents the model without
  merging (Section \ref{sec:model1:form}); the dashed line represents
  the model with merging, only considering the main progenitors
  (Section \ref{sec:model2:main}); and the solid line represents the
  standard version of our model, including merging and considering all
  progenitors (Section \ref{sec:model2:accr}). Our dry-merging model
  successfully reproduces both size and mass density evolution from
  $z=2$ to the present.}
\label{evolplot}
\end{figure}

The population of smaller progenitors by definition have smaller
masses than the population of main progenitors. Their inclusion will
mostly affect the predicted comoving mass density evolution at $z=2$
(see Figure \ref{MR}).  Quantitatively, the model predicts
$\delta\rho(1)=0.45$ and $\delta\rho(2)=0.05$, both consistent with
the data (see Figure \ref{evolplot}). The effect on the predicted
evolution in average size at fixed mass is less pronounced: the model
predicts $\delta\reff(1)=0.63$ and $\delta\reff(2)=0.40$, not much
slower than observed (the remaining disagreement is of order
$1-2\sigma$).

Overall, it is quite remarkable that such a simplistic estimate of the
redshift at which galaxies transform into early types and the simple
assumptions regarding their subsequent merger activity can come so
close to reproducing the observations.

\section{MODEL UNCERTAINTIES}\label{sec:discussion}
The model presented in Sections \ref{sec:model1} and \ref{sec:model2}
is a drastically simplified description of the formation and evolution
of early-type galaxies. In this section we describe some of the
uncertainties that arise from the simplifications and their potential
impact on the model predictions.

\subsection{Uncertainties in the Estimate of the Formation Epoch}\label{sec:discussion:1}
As stated in Section \ref{sec:model1:form}, the value of the exponent
in Equations \ref{sigz} and \ref{zsig} was chosen \textit{a
  posteriori}, $\alpha=3/4$ producing the best results.  The model
predictions are quite sensitive to the choice of $\alpha$. Trying a
range of values for $\alpha$, we find that, for the 'standard' version
of our model as described in Section \ref{sec:model2}, in order to be
consistent with all data at the 2$\sigma$ level, the exponent has to
have a value in the range $0.65<\alpha<0.85$. This is consistent with
the results of \citet{franx08}.  For lower values of $\alpha$ we find,
in particular, that the predicted evolution of the mean size is slower
than observed.  For higher values the predicted comoving mass density
at $z\sim 2$ is lower than observed.  Clearly, if we change the
prescription for merging (see Sections \ref{sec:discussion:2} and
\ref{sec:discussion:3}), the allowed range for $\alpha$ also changes;
however, in order for the model to remain consistent with the observed
evolution of the comoving mass density, $\alpha$ has to be close to
0.75.

A related problem is a possible systematic uncertainty in the velocity
dispersions of nearby early-type galaxies. This remains to be an
issue, especially for the most massive galaxies \citep{bernardi07a}.
Even a 5\% systematic error has major consequences for the $z=2$
predictions because the number of galaxies with high dispersions
declines very rapidly with increasing dispersion.  This implies that
the formation-redshift distribution is rapidly, and perhaps
artificially, truncated as well.  Increasing all velocity dispersions
by 5\% leads to a higher comoving mass density at $z=2$ by as much as
a factor of 3 (while the increase at $z=1$ is just 10\%), whereas size
evolution is only slowed down by less than 5\% over the entire
redshift range.  We conclude that the predictive power of our model
regarding the comoving mass density evolution at $z\sim 2$ is likely
limited by systematic uncertainties in the velocity dispersions of
massive nearby galaxies.

A further simplification in our model is that there is no scatter in
the relationship between velocity dispersion and formation
redshift. As a result, the predicted comoving mass density evolution
will be overestimated, particularly at $z\sim 2$ where the model
predictions depend strongly on the steepness of the mass function. It
is beyond the scope of this paper to properly implement such scatter,
as this would require a more advanced synthesis between the formation
epoch and merger history of the galaxies in our simulated samples.

Finally, we assume that the truncation of star formation coincides
with a morphological transformation. Even though those events may have
the same cause, they need not occur simultaneously.

\subsection{Uncertainties in the Estimated Merger Histories}\label{sec:discussion:2}
We make the approximation (see Section \ref{sec:model2:history}) that
the merging of dark matter halos is followed by the merging of the
galaxies within $\Delta t<<t_{\rm{Hubble}}$ due to dynamical friction.
In reality, galaxies occupying merging halos may merge not at all,
such as in the case of infall of a galaxy-sized halo into a
cluster-sized halo, or they may merge quickly, such as in the case of
two galaxy-sized, equal-mass halos. Generally speaking, the larger the
mass-ratio of a halo-halo merger, the longer the delay of the
galaxy-galaxy merger \citep[see, e.g.,][]{boylan08, kitzbichler08}.
It may, therefore, be necessary to invoke a delay for galaxy mergers
with respect to halo mergers. To obtain an upper limit, we test the
effect of a long, 2.5 Gyr delay on our model predictions, which is of
the same order as the Hubble time at high redshift, and amounts to a
few times the dynamical time scale of large clusters.  In practice,
this means that the galaxy merger activity between $z=1$ and the
present reflects halo merger activity between $z=2$ and $z=0.2$. As a
consequence, major merger activity (with mass ratios of 1:2 or less)
is boosted substantially from 0.25 to 0.5 mergers per galaxy between
$z=1$ and the present.  This results in the stronger size evolution
($\delta\reff(1)=0.57$ and $\delta\reff(2)=0.35$) which would be in
excellent agreement with the data (see Figure \ref{evolplot}).  Both
the observational and model uncertainties are too large, however, to
draw the conclusion that a delay in galaxy merger activity is
preferable, especially because the mass dependence of this phenomenon
is not taken into account.

A merger delay for the galaxies under consideration may indeed not be
of great importance, as cosmological simulations that explicitly treat
galaxy mergers as separate from halo mergers find similar merger rates
for galaxies as we find for halos \citep{maller06,khochfar08}.  This,
along with our finding that any merger delay only increases the
redshift evolution of the mean sizes (i.e., our model in this sense
provides a lower limit), implies that our conclusions are not
compromised by the simple assumption that galaxies mergers are
instantaneous reflections of halo mergers.

\subsection{Uncertainties in the Properties of Progenitors}\label{sec:discussion:3}
Our modeling assumes that in a dry merger, the size of the main
progenitor grows proportionally to its mass increase.  We justify this
choice in Section \ref{sec:model2}, but it is useful to discuss its
effect on the model predictions.  If a substantial amount of orbital
energy were transferred from the stellar component to the dark halo,
the descendant would be more tightly bound than the progenitors. This
would result in a fractional size change less than that in mass, which
is what \citet{boylan05} find in their simulations. The results from
\citet{boylan05} in fact suggest that dry mergers move galaxies along
the slope of the present-day mass-size relation, in which case no
redshift evolution in size at fixed mass due to merging would be
expected. Our approach implies that no energy is transferred to the
halo, resulting in large, relatively loosely bound merger remnants
(with an unchanging velocity dispersion), in agreement with most other
simulations (see Section \ref{sec:model2:mergesize}).  This is why
mergers play an important role in shaping the early-type galaxy
population in our modeling. Conversely, we would infer that the strong
redshift evolution of the mean size implies that the amount of energy
transferred to the halo must be small.

The above arguments make it very clear that our model would not work
without the assumption that galaxies, at least the main progenitors,
grow in size more or less proportional to their mass. The properties
of the smaller progenitors before merging with the main progenitors
are a separate, but related issue. In Section \ref{sec:model2:accr} we
justified our assumption that all progenitors partaking in a merger
have nearly the same velocity dispersion. This breaks down at lower
redshifts, where a reservoir of galaxies with a large range in mass
(and dispersion) is available for merging. It is therefore useful to
test the sensitivity of the model predictions on this assumption. To
do so, we consider the following alternative scenario: the smaller
progenitors follow the present-day mass-size relation with respect to
the main progenitor, i.e., we assume that their size ratios are
$(M_{\rm{accr}}/M_{\rm{main}})^{0.56}$ \citep{shen03,vanderwel08c}.
The virial theorem requires lower velocity dispersions for the smaller
progenitors than for the main progenitor.  By comparing this lower
velocity dispersion with the threshold associated with the particular
redshift that is being simulated (i.e, $z=1$ or $z=2$), we can decide
in a very natural manner within the context of our model whether the
smaller progenitor is a late-type galaxy or an early-type galaxy. If
it has such a low dispersion that the implied formation redshift
according to Equation \ref{zsig} is lower than the simulated redshift,
it is omitted from the sample, otherwise it is retained. Despite the
substantially different approach, this exercise has a limited effect
on the model predictions.  The predicted size evolution is $\sim 10\%$
slower, and the predicted comoving mass density evolution $\sim10\%$
faster.  We conclude that the assumptions regarding the properties of
the smaller progenitors are not of crucial importance to the model
predictions.

It is, in principle, possible that in a (minor) merger the relative
increase in size is larger than the relative increase in mass.  In the
case of virialized progenitors and remnants, homology, and zero energy
transfer between the stars and the dark matter halo, following the
analytical work of \citet{boylan05}, we derive that the ratio of the
radii of the merger remnant ($R_r$) and the main progenitor ($R_m$) is

\begin{equation}
\label{size}
  \frac{R_r}{R_m}=\frac{(1+n)^2}{n^\alpha+n^2},
\end{equation}

where $n$, as before, is the mass ratio of the main progenitor and the
accreted progenitor, and $\alpha$ is the slope of the mass-size
relation for the progenitors ($R\propto M^\alpha$).  For
$n\rightarrow\infty$, we have that $R_r/R_m\rightarrow 1+2/n$ (for
$\alpha<1$), implying that for minor accretion events, the relative
increase in size is close to twice the increase in mass \citep[see
also,][]{naab09}.  This may provide an explanation for the evolution
of the most compact galaxies in the observed $z>2$ samples
\citep[e.g.,][]{zirm07,vandokkum08b}, which our model does not
reproduce.  The difficulty in the context of our model, however, is
that knowledge of the nature of very small accreted galaxies, which in
this scenario would be the main contributors to size evolution,
becomes essential.  It seems unrealistic to suppose that low-mass
galaxies are not gas rich and are not forming stars.  As a
consequence, mergers will be dissipative, with a different effect on
the evolution of the scaling relations \citep[see,
e.g.,][]{robertson06,ciotti07,hopkins09a}.  A more complete
description of the merger history of galaxies, including a
prescription for dissipation, will be required to fully explore the
role of mergers with large mass ratios.  However, if energy transfer
to the halo is a minor factor, the large number of expected minor
accretion events could be a viable explanation for the most compact,
observed galaxies at $z>2$.

If minor mergers that are gas poor dominate the growth of early-type
galaxies, size evolution can be regarded as the assembly of a
low-density envelope around a high-density central region. In such a
scenario, the compact, high-redshift galaxies survive as the centers
of present-day, massive galaxies. The properties of the centers of
local early-type galaxies and the distant compact galaxies are,
indeed, not significantly different \citep{bezanson09, hopkins09b}.
On the other hand, there is no indication that the outer envelopes of
early-type galaxies, the wings of De Vaucouleurs-type surface
brightness profiles, have evolved since $z\sim 1$
\citep{vanderwel08c}.  We conclude that it is thus far unclear, from
an empirical perspective, what the relative contributions of major and
minor merging are.

Summarizing, there are two elements of the model that are essential to
any success in explaining the observations: first, the evolving
velocity dispersion threshold which is associated with the formation
of early-type galaxies; second, the growth of the main progenitor of
present-day early-type galaxies by means of the accretion of mass
through dry mergers in such a manner that size grows proportionally to
mass.

\section{TESTABLE PREDICTIONS}\label{sec:discussion:4}

Our model predicts mass-dependent size evolution, with the high-mass
end of the mass-size relation changing least with redshift (see Figure
\ref{MR}).  This is mainly due to the higher formation redshifts of
more massive galaxies. Such a trend is furthermore strengthened by the
presumed properties of the accreted progenitors described in
Section \ref{sec:model2:accr}. This prediction is exactly opposite to the
prediction by \citet{khochfar06b} that massive galaxies will display
the strongest evolution. This seeming discrepancy may be explained if
the star-formation histories of galaxies in the \citet{khochfar06b}
model are closely linked to the assembly histories of their halos.  We
explicitly separate these processes by invoking a criterion for
galaxies transforming into early types that is unrelated to the
assembly history, only considering further assembly through dry merger
events after the epoch of their initial emergence as early-type
galaxies.

Unfortunately, the current observations are not of sufficient quality
to decide this matter.  The $z\sim 1$ mass-radius relation of the
sample of early-type galaxies with dynamical mass measurements used by
\citet{vanderwel08c} has a somewhat steeper slope than the local
mass-size relation \citep[see also][]{ferreras09}, but this is a
marginal effect (of order $2\sigma$), and which, in addition, could
also be explained as a difference between cluster and field galaxies.
Interestingly, the $z\sim 2$ samples show a hint that the reverse may
be the case \citep{vanderwel08c}, with samples with the largest masses
showing the largest offsets from the local mass-size relation, in
qualitative (but not quantitative) agreement with the predictions by
\citet{khochfar06b}.

Another feature predicted by our model is a decrease with redshift in
the scatter in size at fixed mass (see Figure \ref{MR}) .  This is
simply because at fixed mass the larger galaxies are younger: at
earlier epochs the larger galaxies at a given mass were not yet
members of the early-type galaxy population, such that the scatter was
smaller than it is nowadays.  How rapidly the scatter is predicted to
evolve with redshift depends on the implementation of merger activity,
but regardless of the details a decrease with redshift is
expected. Therefore, a measurement of the scatter in the high-redshift
mass-size relation mainly tests our idea that there is a relation
between age and size at fixed mass, which we already know to exist
given the archaeological properties of present-day early-type galaxies
(see Figures \ref{R_AGE1} and \ref{R_AGE}).  The effect of dry mergers
on the scatter is, in our model, of secondary importance. The $z\sim
1$ observations are consistent with a non-evolving scatter (within
$2\sigma$), but the best-fitting value points toward smaller scatter
at earlier epochs \citep[see][]{vanderwel08c}, in agreement with our
model. Clearly, larger samples are required to confirm this tentative
result.

Our study here concerns several fundamental scaling relations for
early-type galaxies.  It is therefore interesting to consider the
implications of our model for another fundamental scaling relation
that we have not mentioned so far: the connection between
super-massive black holes and their host galaxies
\citep[e.g.,][]{kormendy95,magorrian98,ferrarese00,gebhardt00,graham01}.
In our dry-merger scenario, the growth of galaxies is obviously
connected to the growth of the mass of the central black hole
($M_{\bullet}$).  We assume that $\sigma_*$ remains constant in dry
mergers (Section \ref{sec:model2:mergesize}), whereas the mass of the
halo, the galaxy, and that of the black hole grow more or less
proportionally.  Hence, the evolution of $M_{\bullet}$ is more closely
linked to the evolution of the total mass of both the halo and the
galaxy than to $\sigma_*$.  However, note that this is not an argument
against a fundamental relationship between $\sigma_*$ and
$M_{\bullet}$ that may have determined the initial shape of the
scaling relations between black-hole and bulge properties.  It is as
yet unclear what the intrinsic scatter in these scaling relations is,
or whether one of the relations is fundamental in the sense that the
intrinsic scatter is zero \citep[e.g.,][]{tremaine02,ferrarese05}.
Nor is it clear if and in which direction the $M_{\bullet}$ scaling
relations evolve with redshift for non-active galaxies, despite recent
progress for galaxies with active nuclei
\citep[e.g.,][]{salviander07,woo08}.  In the context of our model,
evolution with redshift is expected, and the intrinsic scatter is
unlikely to be zero.  Scatter and redshift evolution in the
$M_{\bullet}$-$\sigma_*$ relation is expected due to continuous dry
merging and the variety of the merger history of galaxies with a given
$\sigma_*$ (see Section \ref{sec:model2:history}). The scatter should
then correlate with the size of the galaxy (bulge).  Furthermore,
assuming that $\sigma_*$ and $M_{\bullet}$ are initially related,
scatter and evolution in the $M_{\bullet}$-$M_{\rm{dyn}}$ (bulge mass)
relation is expected due to the $\sigma_*$-dependent formation
redshift of galaxies, and the scatter in the relationship between
$\sigma_*$ and $M_{\rm{dyn}}$ (see Section \ref{sec:model1:form}).
The scatter should then correlate with size and the age of the stellar
population.  However, the current observational constraints on both
the scatter of the present-day black-hole scaling relations and their
evolution with redshift are not sufficiently accurate to provide
evidence against or in favor of our model.  Indications that the
scatter indeed correlates with a third parameter have been found
\citep[e.g.,][]{marconi03}, but the statistical significance of these
results is quite marginal and their interpretation still debated
\citep[e.g.,][]{graham08}.

In addition to the $M_{\bullet}$-$\sigma_*$ relation, the model
presented here also predicts some characteristics of other scaling
relations, such as the Faber-Jackson \citep{faber76} and
$Mg_2-\sigma_*$ \citep{dressler87} relations.  Equal-mass, dry mergers
have little effect on the $Mg_2-\sigma_*$ relation, as both the
metallicity and the velocity dispersion remain unchanged.  The scatter
may decrease somewhat, as merging will lead to regression to the mean
metallicity at fixed velocity dispersion.  Unequal-mass mergers will
slightly decrease the metallicity of the merger remnant compared to
the metallicity of the main progenitor (because of the
mass-metallicity relation), which may somewhat increase the scatter in
the $Mg_2-\sigma_*$ relation.  If we assume that luminosity is a proxy
for mass, our model predicts scatter in the Faber-Jackson relation.
This reflects the variety in the merger history of galaxies and is
equivalent to the expected scatter in the $M_{\bullet}$-$\sigma_*$
relation.  Hence, our model predicts a correlation between the scatter
in the $M_{\bullet}$-$\sigma_*$ relation and the Faber-Jackson.  Both
of these should also correlate with galaxy size.  The correlation
between the scatter in Faber-Jackson relation and galaxy size is, of
course, well known.  The existence of the fundamental plane is,
therefore, consistent with a scenario in which dissipationless merging
plays an important role.

\section{SUMMARY AND CONCLUSIONS}
We present a simple, empirically motivated model that simultaneously
predicts the evolution in the mean size and the comoving mass density
of early-type galaxies between $z=2$ and the present.  A large sample
of nearby early-type galaxies extracted from the Sloan Digital Sky
Survey serves as the starting point. The redshifts at which galaxies
transform into early-type galaxies (i.e., their 'formation' redshifts)
are estimated based on an evolving velocity dispersion threshold,
$\sigma_{\rm{ET}}$, above which galaxies have low specific
star-formation rates.  This is motivated in part by the observation
that the stellar population age of present-day early-type galaxies is
a simple function of their velocity dispersion.  In addition to a
prescription for estimating formation epochs, several merging
scenarios between the formation epoch and the present are explored.
Merger trees are generated such that they match the results from
simulations of the mass assembly history of dark matter halos.
Assuming dissipationless ('dry') merging and a simulation-inspired
prescription for the effect of such mergers on the size of galaxies,
we successfully reproduce both the observed evolution in the mean size
and in comoving mass density of early-type galaxies with mass $M >
10^{11}~\msol$ between $z=2$ and the present. Our model quantitatively
explains the recently measured, substantial size evolution of
early-type galaxies. Our model does so by combining the continuous
emergence of early-type galaxies, starting with systems with the
highest velocity dispersion, with subsequent merging.  Recently, and
from the very different perspective of hydrodynamical simulations,
\citet{hopkins09a} also considered such a combination of growth of the
population and growth of individual galaxies, and arrived at similar
conclusions.

We emphasize that our model for estimating the 'formation' redshifts
of early-type galaxies is entirely phenomenological.  We have, so far,
not addressed the question why the velocity dispersion would be
closely related to the formation redshift.  Surely, this is related to
the formation of bulges, perhaps regulated by feedback mechanisms. An
obvious connection can be made between the crucial role that velocity
dispersion plays in our model and the importance of black hole
growth. It will be interesting to see whether a theoretical framework
can be constructed that explains from first principles that early-type
galaxies which form during a certain epoch have velocity dispersions
that are independent of mass.

The most troubling issue of our model, which affects the merger part,
is the effect of mergers on the sizes of the descendants. As mentioned
in Section \ref{sec:model2:mergesize}, some confusion remains in the
literature describing the results from numerical simulations.  Even if
this issue is resolved, it will remain unclear how a multitude of
minor accretion events will affect size evolution.  In principle, a
stronger effect on the sizes of the main progenitors than we assume in
our model is physically possible.  This may explain the presence of
very compact galaxies in the $z>2$ samples; however, dissipation will
have to be incorporated in our model to make realistic predictions,
which is beyond the scope of this study.

Besides the possibly underestimated effect of minor accretion events
in our model, there are other potential solutions to the puzzle that
the most compact galaxies present.  First, the systematic
uncertainties in the observations may still be underestimated; the
ultimate test would be a direct measurement of the velocity
dispersion.  Initial steps in this direction were recently taken by
\citet{cenarro09}.  They find, based on an analysis of stacked
spectra, that the velocity dispersions of quiescent galaxies at
$z\sim1.5$ may be somewhat higher than those of present-day galaxies
with the same stellar mass.  Second, physical mechanisms besides
merging may play a role.  A recently proposed idea is that quasar
feedback is perhaps responsible for the observed size evolution,
through the ejection of gas, simultaneously providing an explanation
for the cessation of star formation, and the increase in size
\citep{fan08}.  An indication that this cannot be the mechanism that
fully explains the observed size evolution is that a substantial
fraction of this evolution took place between $z=1$ and the present;
over this period, massive early-type galaxies are known not to undergo
phases of intensive nuclear activity. On the other hand, quasar
feedback could contribute at higher redshifts, providing an
explanation for the remaining discrepancy between our model
predictions and the most compact galaxies found at $z>2$.

\acknowledgements{We thank the referee, Luca Ciotti, for useful
  suggestions.  A.~v.~.d.~W.~is grateful to Genevieve Graves for
  sharing her SDSS early-type galaxy sample, and acknowledges
  interesting discussions with and useful comments from Marijn Franx,
  Ryan Quadri, Aday Robaina, Andrew Zirm, and Brad Holden. Funding for
  the SDSS and SDSS-II has been provided by the Alfred P. Sloan
  Foundation, the Participating Institutions, the National Science
  Foundation, the U.S. Department of Energy, the National Aeronautics
  and Space Administration, the Japanese Monbukagakusho, the Max
  Planck Society, and the Higher Education Funding Council for
  England. The SDSS Web Site is http://www.sdss.org/.  The SDSS is
  managed by the Astrophysical Research Consortium for the
  Participating Institutions. The Participating Institutions are the
  American Museum of Natural History, Astrophysical Institute Potsdam,
  University of Basel, University of Cambridge, Case Western Reserve
  University, University of Chicago, Drexel University, Fermilab, the
  Institute for Advanced Study, the Japan Participation Group, Johns
  Hopkins University, the Joint Institute for Nuclear Astrophysics,
  the Kavli Institute for Particle Astrophysics and Cosmology, the
  Korean Scientist Group, the Chinese Academy of Sciences (LAMOST),
  Los Alamos National Laboratory, the Max-Planck-Institute for
  Astronomy (MPIA), the Max-Planck-Institute for Astrophysics (MPA),
  New Mexico State University, Ohio State University, University of
  Pittsburgh, University of Portsmouth, Princeton University, the
  United States Naval Observatory, and the University of Washington. }


\bibliographystyle{apj}

\end{document}